# Fast and high-accuracy measuring technique for transmittance spectrum in VIS-NIR

WANG Sheng-hao, LIU Shi-jie, WANG Wei-wei, ZHANG Zhi-gang

Testing center, Shanghai Institute of Optics and Fine Mechanics, Chinese Academy of Sciences, Shanghai, 201800, China.

Email address: wangshenghao@siom.ac.cn, shijieliu@ siom.ac.cn

**Abstract:** In this paper, based on the framework of traditional spectrophotometry, we put forward a novel fast and high-accuracy technique for measuring transmittance spectrum in VIS-NIR wave range, its key feature is that during the measurement procedure, the output wavelength of the grating monochromator is kept increasing continuously and at the same time, the photoelectric detectors execute a concurrently continuous data acquisition routine. Initial experiment result shows that the newly proposed technique could shorten the time consumed for measuring the transmittance spectrum down to 50% that of the conventional spectrophotometric method, a relative error of 0.070% and a repeatability error of 0.042% are generated. Compared with the current mostly used techniques (spectrophotometry, methods based on multi-channel spectrometer and strategy using Fourier transform spectrometer) for obtaining transmittance spectrum in VIS-NIR, the new strategy has at all once the following advantages, firstly the measuring speed could be greatly quicken, fast measurement of transmittance spectrum in VIS-NIR is therefore promising, which would find wide application in dynamic environment, secondly high measuring accuracy (0.1%-0.3%) is available, and finally the measuring system has high mechanical stability because the motor of the grating monochromator is rotating continuously during the measurement.

**Keywords:** Fast, High-accuracy, VIS-NIR, Transmittance spectrum, Measurement

## 1. Introduction

As a strong and powerful characterization means, transmittance spectrum in the wave range of VIS-NIR has been used widely not only in scientific research fields such as physics, chemistry, biology, but also in chemical engineering, medicine and many other industrial areas [1-4]. Currently the mostly popular techniques for measuring transmittance spectrum in VIS-NIR are spectrophotometry [5-10], method based on multi-channel spectrometer [11-13] and strategy using Fourier transform spectrometer [14-15].

In the measuring technique of spectrophotometry, typically a prism [5-6, 9] or a diffraction grating [7-8, 10] is adopted as the dispersing element in order to generate the monochromatic light beam, and during the testing procedure, firstly the transmittance rate of the sample is measured and computed at the wavelength corresponding to the current monochromatic beam with the help of photoelectric detectors, after that by rotating the dispersing element and repeating the aforementioned process, the transmittance rate at each individual wavelength in the expected wave range can be obtained, and finally in this way, transmittance spectrum of the inspected sample within a certain wave range can be acquired. Spectrophotometry has very high measuring accuracy (about 0.1%-0.3%), but the slow measuring speed resulting from the hundreds of repetitive actions at each wavelength is its significant drawback, for example, about 3-5 minutes are needed to accomplish the measurement of transmittance spectrum in the wave range of 500-900 nm [12-13], meanwhile, considering the output wavelength of the monochromator is changing step by step through rotating the dispersing element during the measurement, the mechanical module of the system has to be kept in a discontinuous movement, the testing equipment thus has very poor mechanical stability. For the measuring method based on multi-channel spectrometer, the polychromatic light beam firstly penetrates the sample and is then diffracted by the grating, after that a linear CCD detector is used to obtain the whole spectrum, based on the mathematic relationship between the wavelength and the pixel position of the linear CCD detector (usually acquired by a calibrating optical source), the transmittance spectrum in the whole expected wave range can be attainable by a single shot strategy. Obviously key advantage of the measuring method based on multi-channel spectrometer is that the spectrum measurement can be accomplished fastly in the time scale of micron second, however the low measuring accuracy (around 5% - 8%) and the inconvenient wave range tuning (it is usually fixed) greatly limits its application [11-13]. About the methodology using Fourier transform spectrometer, the transmittance spectrum is obtained by Fourier transform upon the interferogram gained from a Michelson interferometer, in this manner the transmittance spectrum can be generated in a few seconds, but just like the aforementioned measuring technology using multi-channel spectrometer, the low accuracy (around 2% - 3%) is its primary defect [14-15].

Spectrophotometry usually finds application in the measuring environment requiring high-accuracy and low time resolution, for example, when charactering optical film, solution and many other types of samples possessing stable optical properties. On the contrary, methods based on multi-channel spectrometer and strategy using Fourier transform spectrometer always get used in the circumstance where high time resolution and low accuracy are demanded, such as online monitoring the chemical reaction, analyzing the thermodynamic process, and many other dynamic charactering applications. However, in the condition where both high accuracy and high time resolution are required, for example, in order to observing with high accuracy the dynamic evolution process of the sample's transmittance spectrum with the change of external temperature, humidity or chemical reaction, the transmittance spectrum is supposed to be obtained in a very short time and with a very high accuracy, and it is apparent that the currently most used techniques are not competent for these applications.

In this manuscript, based on the hardware configuration of traditional spectrophotometry, and different from the traditional serial data collection mode (wavelength change → data acquisition → wavelength change → data acquisition → ⋯), we propose a parallel data acquisition strategy, in which the output wavelength of the grating monochromator is kept increasing continuously and at the same time, the photoelectric detectors execute a concurrently continuous data acquisition routine, our work is aimed at primarily evaluating the new fast and high-accuracy technique for measuring transmittance spectrum in VIS-NIR wave range.

## 2. Materials and methods

### 2.1 Transmittance spectrum measuring equipment

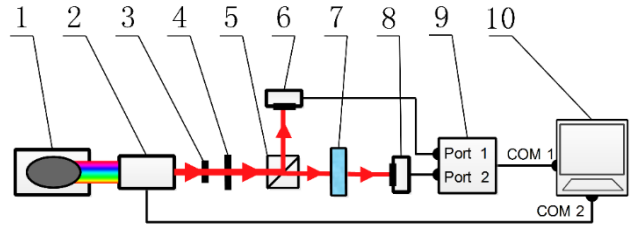

Fig. 1 Framework of the system for measuring transmittance spectrum in VIS-NIR wave range, 1- supercontinuum laser source, 2- grating monochromator, 3- aperture, 4- line polaroid, 5- unpolarized beam splitter, 6-referencing detector, 7-sample, 8-testing detector, 9-data acquisition module, 10-personal computer.

As shown in Fig. 1 is the framework of the system for measuring transmittance spectrum in VIS-NIR wave range, it is mainly made up of a supercontinuum laser source, a grating monochromator, an aperture, a line polaroid, an unpolarized beam splitter, a referencing detector, a testing detector, a set of data acquisition module and a personal computer. The supercontinuum laser source (Manufactory: Fianium, Model: TLSS-VIS-HP-2) emits a polychromatic spectrum with a total power of 2 W in 390-2600 nm wave range, and the mean power density is about 1 mw/nm, while its power stability is better than ±1%. The grating monochromator (Manufactory: Photon, Model: CONTRAST X-HP4) can operate from 500 nm to 1200 nm with a wavelength resolution around 0.125 - 0.3125 nm, and the FWHM of the produced light beam from the monochromator is about 1.0 - 2.5 nm, when the output wavelength of the monochromator is changing in the step of 0.1 nm, a stabilization time of at least 35 microsecond is needed, while it is 55 microsecond for the increasing step of 1 nm. The line polaroid (Manufactory: Thorlabs, Model: LPVIS050-MP2) has an effective wave range of 550 - 1500 nm, and its extinction ratio is higher than 1000:1. The unpolarized beam splitter (Manufactory: Thorlabs, Model: CM1-BS014) can be applied from 700 nm to 1100 nm with a splitting ratio of 50:50. The referencing photoelectric detector and the testing detector (Manufactory: OPHIR, Model: PD300-1W P/N 7Z02411A) are photodiode-based power meter, their spectral response wave range is 350 - 1100 nm, and the power detecting range is between 500 pw and 1 w. The sample under inspection is a silica glass coated by a 500 nm thick photoresist (Manufactory: Shimpy, Model: S1805) film.

The polychromatic light beam emitted by the supercontinuum laser source is first transmitted by optical fiber into the grating monochromator, and then the generated monochromatic beam passes successively through the optical aperture and the line polaroid, by now the expected high quality polarized beam is produced, after that a referencing light beam and a testing beam are simultaneously generated under the action of the unpolarized beam splitter, the referencing photoelectric detector works here for collecting the referencing light beam, while the testing detector captures the testing beam after it penetrating the inspected sample, the photo signal detected by the two detectors is firstly photovoltaic transformed and then digitized by the dual channel data acquisition module. The personal computer is used to remote control the grating monochromator and operate the data acquisition module through serial ports, a LabVIEW-based software platform is developed to implement the desired functionality such as system initialization, wavelength setting, wavelength scanning, data acquisition, data post processing, graphic display and data storage.

**2.2 Theory and procedure of the traditional Spectrophotometry**

Based on the measuring framework as shown in Fig.1, theory and procedure of the traditional spectrophotometry for obtaining transmittance spectrum in VIS-NIR is written as following:

S1: Remove the sample off the testing light beam, and set the output wavelength of the monochromator at $\lambda_1$, then capture simultaneously the power of the referencing light beam and the testing beam through the dual channel data acquisition module, write respectively as $I_1(\lambda_1)$ and $I_2(\lambda_1)$, compute and record the current intensity ratio $k(\lambda_1)$ as:

$$k(\lambda_1) = \frac{I_2(\lambda_1)}{I_1(\lambda_1)} \tag{1.1}$$

S2: Set successively the output wavelength of the monochromator at $\lambda_2$, $\lambda_3 \cdots \lambda_{n-1}$ and $\lambda_n$, meanwhile repeat step S1 at each individual wavelength, compute and store the intensity ratio $k(\lambda_2)$, $k(\lambda_3) \cdots k(\lambda_{n-1})$ and $k(\lambda_n)$ corresponding to each wavelength.

S3: Place the sample into the testing light beam, and set the output wavelength of the monochromator at $\lambda_1$, then acquire simultaneously the intensity of the referencing light beam and the testing beam, write respectively as $I_1(\lambda_1)$ and $I_2(\lambda_1)$, compute and record the current intensity ratio $k^*(\lambda_1)$ as:

$$k^*(\lambda_1) = \frac{I_2^*(\lambda_1)}{I_1^*(\lambda_1)} \tag{1.2}$$

S4: Set successively the output wavelength of the monochromator at $\lambda_2$, $\lambda_3 \cdots \lambda_{n-1}$ and $\lambda_n$, meanwhile repeat step S3 at each individual wavelength, compute and store the intensity ratio $k^*(\lambda_2)$, $k^*(\lambda_3) \cdots k^*(\lambda_{n-1})$ and $k^*(\lambda_n)$ corresponding to each wavelength.

S5: Use equation (1.3) to calculate respectively the transmittance rate, that is $T(\lambda_1)$, $T(\lambda_2)$, $T(\lambda_3) \cdots T(\lambda_{n-1})$ and $T(\lambda_n)$ of the sample corresponding to the wavelength of $\lambda_1$, $\lambda_2$, $\lambda_3 \cdots \lambda_{n-1}$ and $\lambda_n$.

$$T(\lambda_i) = \frac{k^*(\lambda_i)}{k(\lambda_i)} \tag{1.3}$$

S6: Based on the obtained transmittance ratio, $T(\lambda_1)$, $T(\lambda_2)$, $T(\lambda_3) \cdots T(\lambda_{n-1})$ and $T(\lambda_n)$, plot the transmittance spectrum of the sample.

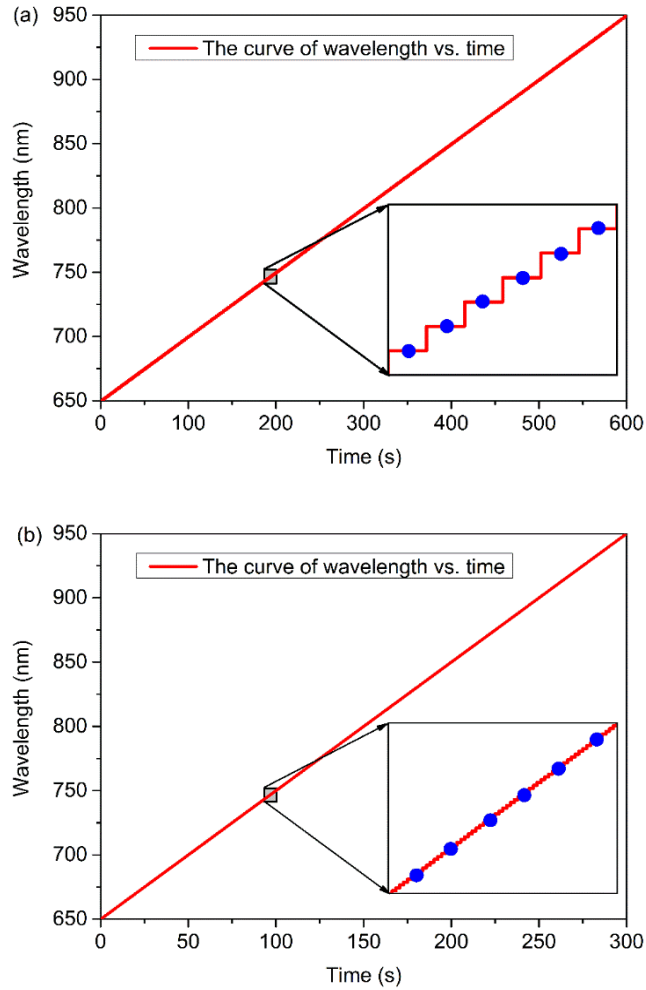

Fig. 2 Schematic diagram of the VIS-NIR transmission spectrum measuring procedure, the red lines in (a) and (b) shows respectively the curves of wavelength vs. time within the traditional and the newly proposed data acquisition strategies, the blue dots in the enlarged part of (a) and (b) represent the moment when the optical detectors work.

Fig. 2(a) demonstrated the testing procedure of the traditional spectrophotometry for the sample in our research, the chosen wave range for measuring transmittance spectrum is 650-950 nm, and the red line in Fig. 2(a) shows the curve of wavelength vs. time during the measurement, we can see from the enlarged view in the bottom right corner that the output wavelength of the grating monochromator is increasing discontinuously step by step, the blue dots represent the moment when the photoelectric detectors capture the signal. In the measurement, firstly the output wavelength of the grating monochromator increases a step of 1 nm, and after a stabilization time of about 0.1 second, the referencing detector and the testing detector collect simultaneously the referencing light beam and the testing beam respectively at the position depicted by the blue dots as shown in Fig. 2(a), then the output of the grating monochromator continues a 1 nm step increasing, and the system keeps repeating the aforementioned process, until all the data is acquired in the whole wave range of 650 - 950 nm, in our experiment about 1 seconds are needed for the system to accomplish the measurement at a single wavelength, and the total time for measuring the transmittance spectrum from 650 nm to 950 nm is about 600 seconds.

**2.3 Testing process of the new method**

Based on the framework as shown in Fig. 1, key feature of the new proposed technique for measuring transmittance spectrum in VIS-NIR is that during the measurement, the output wavelength of the grating monochromator is kept increasing continuously and at the same time, the photoelectric detectors execute a concurrently continuous data acquisition routine. As depicted in Fig. 2(b) is the testing process of the new transmittance spectrum measuring strategy, the red line in Fig. 2(b) shows the curve of wavelength vs. time during the measurement, and the output wavelength is increasing linearly from 650 nm to 950 nm with a changing rate of 1 nm/second, after starting the measurement, the photoelectric detector executes continuous data acquisition at the moment demonstrated by the blue dots as shown in the enlarged view in Fig. 2(b), the time interval between the two adjacent data acquisition position is 0.5 second, and the total time required to accomplish the whole spectrum measurement is about 300 seconds.

It should be pointed out here that the in our current practical experiment the output wavelength of the grating monochromator in fact is not changing continuously as the experimental condition is restricted to the present hardware configuration. In this research the experimental condition that the output wavelength of the grating monochromator should increase continuously is approximately simulated by a " 0.1 nm increasing → 0.1 second pause → 0.1 nm increasing → 0.1 second pause →⋯" manner, we think this approximate treatment is reasonable and feasible because the data acquisition time interval of the photoelectric detector is 0.5 second.

The new measuring strategy holds the similar data post processing procedure with the traditional spectrophotometry when computing the transmittance spectrum, and the difference is that in the new method an additional mathematic process is required to calculate the wavelength corresponding to each raw optical power data, which can be accomplished by combining the initial wavelength, wavelength's changing rate and the sample frequency of the photoelectric detector.

## 3. Experimental results and discussion

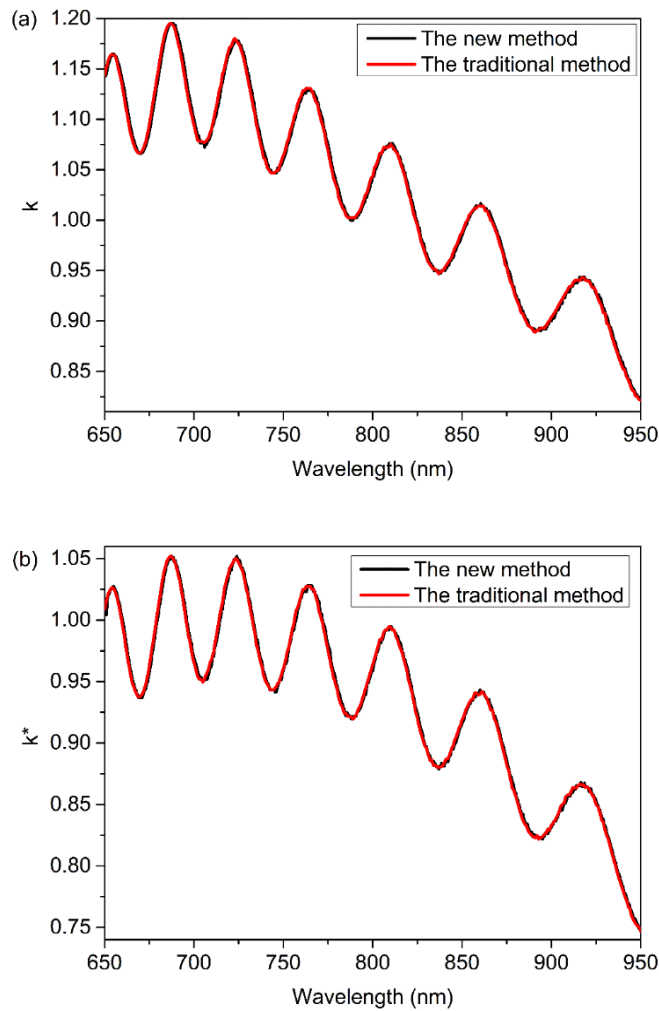

Fig. 3 The measured curves of intensity ratio vs. wavelength, (a) and (b) show respectively the obtained curve without and with sample in the beam path, the red curve represents the measured data using the traditional method while the black one depicts that with the new measuring strategy.

The measured curves of intensity ratio vs. wavelength in the wave range of 650 - 950 nm are shown in Fig. 3, Fig. 3(a) and Fig. 3(b) are respectively the obtained curve without ($k(\lambda)$) and with ($k^*(\lambda)$) sample in the testing beam path, the red curve represents the measured data using the traditional spectrophotometry while the black one depicts that with the new measuring strategy. We can see from Fig. 3(a) and Fig. 3(b) that for the both measured curves, the new proposed measuring strategy has very high consistency with the traditional spectrophotometry.

As demonstrated in Fig. 4 is the computed transmittance spectra in the wave range of 650 - 950 nm from the raw data $k(\lambda)$ and $k^*(\lambda)$, Fig. 4(a) represents comparison of the obtained transmittance spectrum by the new proposed measuring strategy and the traditional spectrophotometry, the black line and the red line depict the transmittance spectra of the traditional method and the new measuring strategy respectively, we want to point out at here that the Savitzky - Golay filter[16] was used in the data post processing to smooth the curves.

The relative error between the two transmittance spectra generated by the new measuring strategy and the traditional method can be written as:

$$\delta = \sqrt{\frac{\sum_{i=650}^{950}\left[\frac{T_{new}(\lambda_i) - T_{trad}(\lambda_i)}{T_{new}(\lambda_i)}\right]^2}{950 - 650}} = 0.070\% \quad (1.4)$$

Where $T_{new}(\lambda_i)$ and $T_{trad}(\lambda_i)$ depict respectively the measured transmittance ratio at wavelength of $\lambda_i$ by the new measuring strategy and the traditional spectrophotometry.

Based on the visual comparison of the two transmittance spectra as shown in Fig. 4(a), and the computed relative error between the two transmittance spectra, we can make a preliminary conclusion that the transmittance spectrum yielded by the new measuring strategy proposed in this manuscript has very high consistency when compared with that of the traditional spectrophotometry.

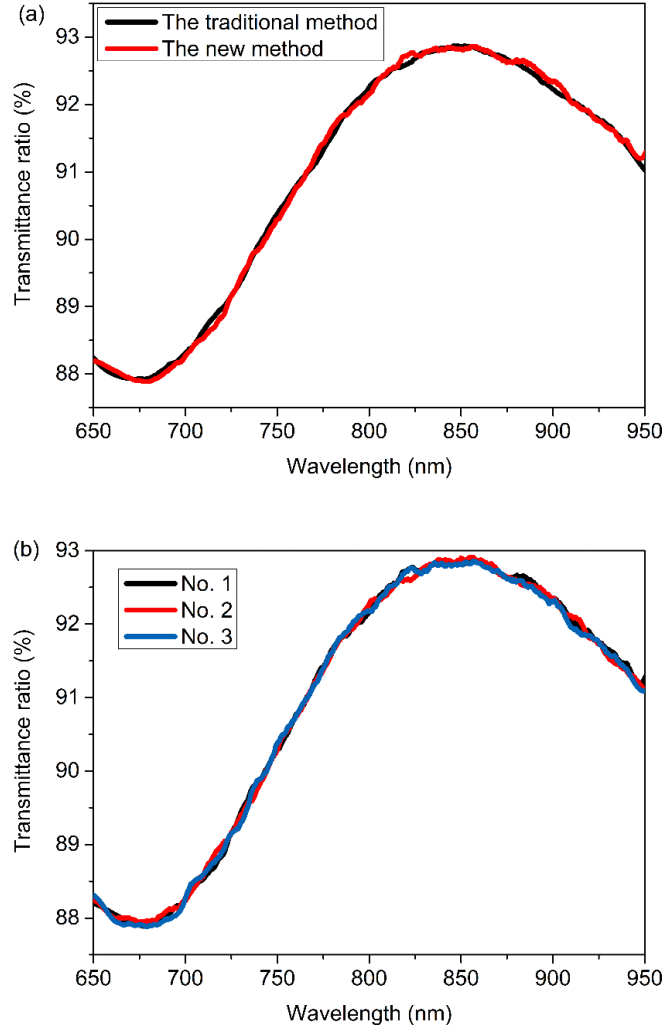

Fig. 4 The measured transmission spectra and the repetitive measurement result of the new method, (a) comparison of the obtained transmittance spectra (black line: traditional method, red line: new method), (b) three times repetitive measurement result using the new method.

Fig. 4(b) represents the three times repetitive experiment result for the transmittance spectra in the range of 650-950nm, and the repeatability error of the three times repetitive experiment can be written as:

$$\zeta = \frac{\sum_{k=650}^{950} \chi_{\lambda_k}}{950-650} = 0.042\% \quad (1.5)$$

Where $\chi_{\lambda_k}$ is the repeatability error at the wavelength of $\lambda_k$, and it was computed as following:

$$\chi_{\lambda_k} = \frac{\sqrt{\dfrac{\sum_{i=1}^{3}\left(T_i - \dfrac{\sum_{i=1}^{3} T_i}{3}\right)^2}{3}}}{\dfrac{\sum_{i=1}^{3} T_i}{3}} \tag{1.6}$$

In equation (1.6), $T_i$ represents the transmittance ratio of the sample at wavelength of $\lambda_k$ in the $i$ th measurement.

Combing the contrast curves as shown in Fig. 4(b) and the aforementioned computed repeatability error (0.042%), it can be concluded that the new measuring strategy has very good repeatability accuracy, taking into consideration that the repeatability error of the traditional spectrophotometry for measuring transmittance ratio is around 0.05%.

## 4. Conclusion and prospection

In conclusion, we put forward a novel fast and high-accuracy technique for measuring transmittance spectrum in VIS-NIR wave range, and its feasibility is preliminarily tested and confirmed by initiatory experimental results. The newly proposed measuring strategy has at all once the following advantages, firstly the measuring speed could be greatly quickened, secondly high measuring accuracy (0.1%-0.3%) is available, and finally the measuring system has high mechanical stability, these key features would make it a novel characterization means in circumstance where high time resolution and high accurate are both required.

As the experimental condition is restricted to the present hardware configuration, the experiment are only carried out in an approximate way, on the one hand, the experimental condition that the output wavelength of the grating monochromator should change continuously is in fact simulated by a " 0.1 nm increase → 0.1 second pause → 0.1 nm increase → 0.1 second pause → ··· manner ", on the other hand, the data acquisition frequency of the photoelectric detector is only 2 times per second. In the future, the system will get upgraded in the following ways, firstly, in cooperation with the manufacturer of the grating monochromator, the output wavelength of the grating monochromator would be supposed to change continuously and fastly for example only 5 seconds is needed for the wavelength increasing linearly from 500 nm to 900 nm, and secondly, high speed photoelectric detector and corresponding data acquisition module would be equipped, systematic experimental research would be carried out by then to evaluate the newly proposed fast and high accuracy transmittance spectrum measuring technique.

**Acknowledgment**

The authors would thank Dr. Fanyu Kong (Key Laboratory of Materials for High-Power Laser, Shanghai Institute of Optics and Fine Mechanics, Chinese Academy of Sciences) for preparing and offering the sample. The research was partly supported by the National Natural Science Foundation of China (No.11602280) and the scientific equipment developing project of the Chinese academy of sciences (No.28201631231100101).